\documentclass{kapproc} 
\usepackage{epsfig}
\kluwerbib

\newcommand{\beq}{\begin{equation}}  
\newcommand{\eeq}{\end{equation}}  
\newcommand{\bea}{\begin{eqnarray}}  
\newcommand{\eea}{\end{eqnarray}}

\begin{document}     
\articletitle{Asymmetric time correlations in turbulent shear flows}
\author{
Bruno Eckhardt, Arne Jachens and J\"org Schumacher
}     
\affil{Fachbereich Physik, Philipps Universit\"at    
	 Marburg, D-35032 Marburg, Germany}    

\begin{abstract}
Cross correlations between normal and downstream velocity
fluctuations in a turbulent shear flow are shown to carry 
information about the non-normal amplification process. 
The creation of spanwise modulated streaks by downstream
vortices implies an asymmetry in temporal correlation functions.
We verify this in numerical simulations in shear flows with
$Re_\lambda\approx 100$. 
\end{abstract}


Coherent structures are an ubiquitous feature of turbulent shear
flows (Townsend 1976, Holmes et al 1996). 
Various kinds of vortices, streaks or waves have been 
identified and considerable efforts have gone into identifying
their dynamical origins and evolution. In boundary layers
the non-normal amplification or lift-up effect 
(Landahl 1980, Boberg and Brosa 1988, Trefethen et al 1992, 
Grossmann 2000) is often an important source for coherent 
structures. Waleffe (1995, 1997) and Hamilton et al (1995) have 
discussed how lift-up and instabilities form a complete
regeneration cycle that can explain sustained large scale
fluctuations. The
aim of our analysis is to find evidence for this process in statistical
measures, in particular in temporal cross-correlation functions.

Following Waleffe (1995, 1997) and Hamilton et al (1995) the
recylcing process has three steps: i) downstream vortices 
mix fluid in the normal direction and drive modulations in the downstream
velocity, forming so-called streaks. ii) streaks undergo an instability 
to the formation of vortices pointing in the normal direction. 
iii) the mean shear profile now turns these vortices again in 
downstream direction, thus closing
the loop. Of these processes the ones in step iii) and ii) are reasonably
fast, whereas the one in i) is fairly slow, since it is connected
with the lift-up and thus only linear in time.
Evidence for this regeneration mechanism
was found in various flows (Hamilton et al 1995, Waleffe 1995, 1997,
Grossmann 2000). Within a dynamical system picture the regeneration 
process can be connected to a periodic orbit, as in the case
of Kawahare and Kida (2001). The complete application of this 
picture to turbulence is complicated not only by the presence 
of many more periodic orbits (as found by Schmiegel (unpublished)
for a low-dimensional model), but also
by the possibility of other spatial variabilities than just a
periodic variation in spanwise and downstream directions. Therefore,
in order to identify this process in fully developed turbulent flows 
other indicators have to be found. 

The indicator for non-normal amplification that we focus on here
is a temporal cross-correlation function (Eckhardt and Pandit 2002). 
Since the vortex drives the streak a cross-correlation between the 
vortex and the streak should be asymmetric in time: if the streak 
is probed after the vortex then there might be a correlation, if 
it is probed before then there should not be a correlation.

The origin of such correlations can be made clear in the context of
a linear analysis around a laminar profile (Eckhardt and Pandit, 2002).
let ${\bf u}_0=Sy{\bf e}_x$ be the shear flow profile,
let $\omega(t)$ the amplitude of a vortex and $s(t)$ be
the amplitude of the streak,
and assume that the nonlinear fluctuations can be modelled 
by white noise. The resulting linear stochastic model can be
solved analytically and the cross-correlation between 
vortex and streak,
\beq
C_{\omega,s}(t) = \langle \omega(t') s(t'+t)\rangle_{t'}\,, 
\eeq
becomes
\beq
C_{\omega,s}(t) = \left\{ \matrix{ -S e^{\lambda t} & t<0 
\cr -S (1+\lambda t) e^{-\lambda t} & t>0} \right. \,.
\eeq
As expected it is asymmetric.
The sign of the correlation function follows from the sign of the 
velocity gradient: if the downstream velocity increases with 
$y$ (positive $S$), then a positive velocity component will bring
up slower fluid, hence make a negative change in the streak. Similarly,
if it brings faster fluid down, it will make a positive contribution
in downstream velocity, but with a negative vertical direction.
So in both cases the cross correlation is negative. In addition,
the correlation function is proportional to the shear rate $S$.
The asymmetry follows from an additional term for positive times,
so that the ratio
\beq
Q(t) = C(t)/C(-t) = 1+ \lambda t
\label{qt}
\eeq
is simply linear.

Clearly, this form of the correlation function is obtained under 
a series of assumptions, some of which are discussed in 
(Eckhardt and Pandit 2002), and further verification is 
required. In boundary layers the mean flow and Taylors frozen
flow hypothesis are usually combined to translate temporal
correlations into spatial correlations (e.g. Townsend 1976). 
This problem is avoided in a Lagrangian frame of reference
or in a comoving frame without mean flow. Evidence
for the asymmetry in a Lagrangian correlation function
can be found in Fig.~1 of (Pope 2002), in a discussion
of stochastic Lagrangian models. For Eulerian correlation
functions we turn to our numerical simulations of a
shear flow (Schumacher and Eckhardt 2000, Schumacher 2001).

The flow is bounded by parallel free-slip surfaces and driven
by a steady volume force that maintains a linear shear profile
(Schumacher and Eckhardt 2000). The statistical properties of 
the flow are in good agreement with other approximations to 
homogeneous shear flows (Schumacher 2001). What is important 
for our analysis is the fact that the mean downstream velocity 
vanishes: we can thus calculate correlation functions in 
a situation without mean flow (in the center) or with 
a slow mean flow (off-center, up to about half the distance 
to the surfaces).

\begin{figure}
\begin{minipage}{0.65\textwidth}
\vskip0.5cm
\epsfxsize=0.9\textwidth
\epsfbox{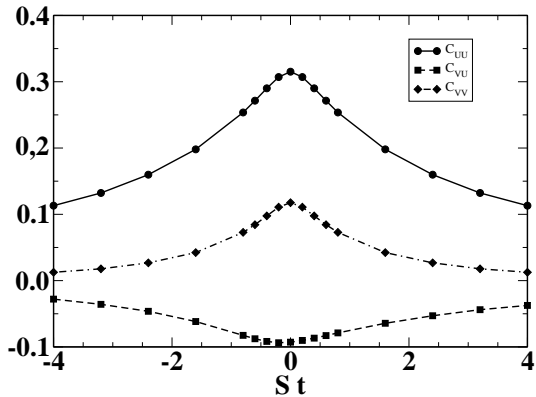}
\end{minipage}\hfill
\begin{minipage}{0.35\textwidth}
\caption[]{Temporal correlation functions in a turbulent shear
flow. The downstream component is $U$, the normal component is $V$.
Times are in dimensionless shear times, the Taylor Reynolds number is 
$Re_\lambda=100$. The auto-correlation functions are symmetric,
the cross-correlation function is negative and slightly
asymmetric.
}
\end{minipage}
\label{correlations}
\end{figure}

\begin{figure}
\begin{minipage}{0.65\textwidth}
\vskip0.5cm
\epsfxsize=0.90\textwidth
\epsfbox{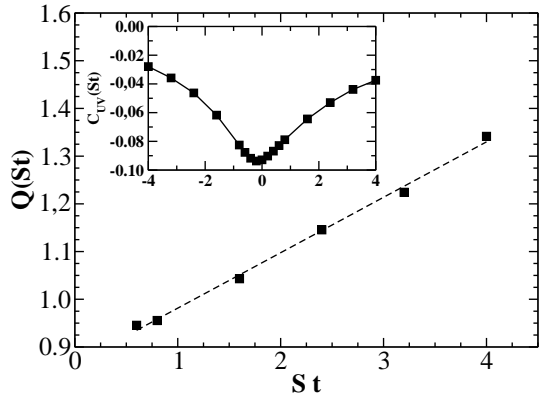}
\end{minipage}\hfill
\begin{minipage}{0.35\textwidth}
\caption[]{Asymmetry in the cross-correlation function for short
times. The large frame shows the ratio of the cross-correlation function
for positive and negative times, $Q(t)=C(t)/C(-t)$, where a linear
behaviour is detected. The inset shows a magnification of the 
cross-correlation function of Fig.~\ref{correlations} near the center.
}
\label{asymmetry}
\end{minipage}
\end{figure}

The results from a simulation with $Re_\lambda\approx 100$
are shown in Fig.~\ref{correlations}. The auto-correlation 
functions of the downstream ($U$) and normal ($V$) velocity components 
are symmetric in time. The cross-correlation function is negative
and slightly asymmetric. In order to analyze the asymmetry further
we show in Fig.~\ref{asymmetry} both a magnification of the 
central region and the ratio (\ref{qt}). The
agreement with the linear prediction is satisfactory,
considering the many non-linear processes in this
fairly turbulent flow.

In summary, we have identified an asymmetry in temporal
cross-correlation functions between downstream and normal
velocity components in turbulent shear flows. The linearity
of the asymmetry supports the connection to the liftup effect.
For the analysis of experimental data the effects of 
a mean flow, of rigid boundaries and also of spatial 
inhomogeneities have to be investigated. We have 
evidence that the effect is strongest near the 
boundary and decreases as one moves
into the turbulent volume. This would be consistent with the
decrease of the mean shear gradient and would indicate 
that non-normal amplification becomes less important
further away from the boundaries. \\

\noindent{\bf Acknowledgements}\\
We thank the Deutsche Forschungsgemeinschaft for support
and the Neumann Center for Computing at the Forschungszentrum
J\"ulich f\"ur computing time and support on their Cray T90.

\end{document}